\documentclass[graybox]{svmult}

\usepackage{amsmath}
\usepackage{mathptmx}			
\usepackage{helvet}         	
\usepackage{courier}        	
\usepackage{type1cm}       		
\usepackage{makeidx}        	
\usepackage{graphicx}       	
\usepackage{multicol}        	
\usepackage[bottom]{footmisc}	
\usepackage{url}			  	

\usepackage{fancyvrb}
\DefineVerbatimEnvironment{code}{Verbatim}{fontsize=\small}
\DefineVerbatimEnvironment{example}{Verbatim}{fontsize=\small}	

\usepackage{listings}
\usepackage{color}

\makeindex             			

\begin{document}

\title*{Data Provenance and Management in Radio Astronomy: A Stream Computing Approach}
\titlerunning{DPDM in Radio Astronomy: A Stream Computing Approach}
\author{Mahmoud S. Mahmoud, Andrew Ensor, Alain Biem, Bruce Elmegreen \newline and Sergei Gulyaev}
\authorrunning{M.S. Mahmoud, A. Ensor, A. Biem, B. Elmegreen \& S. Gulyaev}
\institute{Mahmoud S. Mahmoud \at AUT Institute for Radio Astronomy \& Space Research, Auckland NZ \newline\email{mmahmoud@aut.ac.nz}
\and Andrew Ensor \at AUT Institute for Radio Astronomy \& Space Research, Auckland NZ \newline\email{aensor@aut.ac.nz}
\and Alain Biem \at IBM T J Watson Research Center, Yorktown Heights NY \newline\email{abiem@ibm.com}
\and Bruce Elemgreen \at IBM T J Watson Research Center, Yorktown Heights NY \newline\email{bge@ibm.com}
\and Sergei Gulyaev \at AUT Institute for Radio Astronomy \& Space Research, Auckland NZ \newline\email{sgulyaev@aut.ac.nz}}
\maketitle

\hyphenation{Info-Sphere} 

\abstract*{New approaches for data provenance and data management (DPDM) are required for mega science projects like the Square Kilometer Array, characterized by extremely large data volume and intense data rates, therefore demanding innovative and highly efficient computational paradigms. In this context, we explore a stream-computing approach with the emphasis on the use of accelerators. In particular, we make use of a new generation of high performance stream-based parallelization middleware known as InfoSphere Streams. Its viability for managing and ensuring interoperability and integrity of signal processing data pipelines is demonstrated in radio astronomy.
\newline\indent IBM InfoSphere Streams embraces the stream-computing paradigm. It is a shift from conventional data mining techniques (involving analysis of existing data from databases) towards real-time analytic processing. We discuss using InfoSphere Streams for effective DPDM in radio astronomy and propose a way in which  InfoSphere Streams can be utilized for large antennae arrays. We present a case-study: the InfoSphere Streams implementation of an autocorrelating spectrometer, and using this example we discuss the advantages of the stream-computing approach and the utilization of hardware accelerators.}

\abstract{New approaches for data provenance and data management (DPDM) are required for mega science projects like the Square Kilometer Array, characterized by extremely large data volume and intense data rates, therefore demanding innovative and highly efficient computational paradigms. In this context, we explore a stream-computing approach with the emphasis on the use of accelerators. In particular, we make use of a new generation of high performance stream-based parallelization middleware known as InfoSphere Streams. Its viability for managing and ensuring interoperability and integrity of signal processing data pipelines is demonstrated in radio astronomy.
\newline\indent IBM InfoSphere Streams embraces the stream-computing paradigm. It is a shift from conventional data mining techniques (involving analysis of existing data from databases) towards real-time analytic processing. We discuss using InfoSphere Streams for effective DPDM in radio astronomy and propose a way in which  InfoSphere Streams can be utilized for large antennae arrays. We present a case-study: the InfoSphere Streams implementation of an autocorrelating spectrometer, and using this example we discuss the advantages of the stream-computing approach and the utilization of hardware accelerators.}

\section{Introduction}
\label{sec:Introduction}
Started in the 1930s, radio astronomy has produced some of the greatest discoveries and technology innovations of the 20th century.  One of these innovations -- radio interferometry and aperture synthesis -- was awarded a Nobel Prize for Physics in 1974 (Martin Ryle and Antony Hewish, 1974). An aperture synthesis radio telescope consists of multiple receiving elements in an array that observe the same radiating source(s) simultaneously.
Essentially, an array of radio telescopes is used to emulate a much large telescope with size that of the diameter of the array, enabling a better angular resolution of the radio source(s) to be obtained. While angular resolution is determined by the array diameter, another important characteristic is telescope sensitivity, which is determined by its collecting area. The Square Kilometer Array (SKA) will be an aperture synthesis radio telescope, scheduled for completion in the 2020s, that will combine both factors, resolution and sensitivity. The total SKA collecting area of one square kilometer ($10^6$ m$^2$) will provide sensitivity that is 50-100 times higher than that of the best current radio telescope arrays. Its high angular resolution will be provided by distributing the square kilometer of collecting area into many stations that are spread out on a continental scale (with the baseline between some antennae over $3000$ km).

\begin{figure}[ht]
\centering
\includegraphics[scale=1.0]{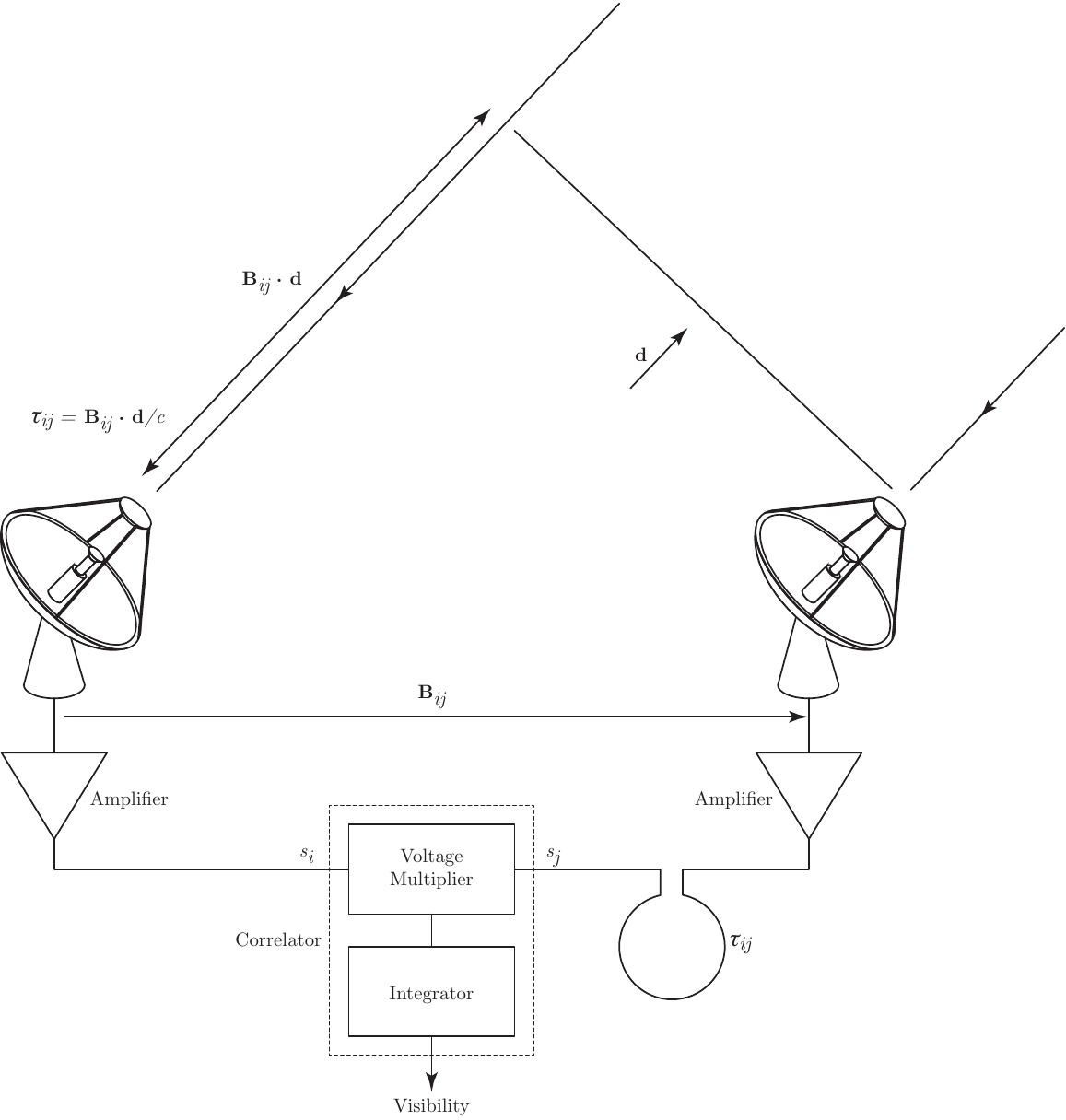}
\caption{Two-element interferometer}
\label{interferometer}
\end{figure}

A radio telescope antenna element detects electromagnetic waves by a current induced in an antenna receiver system. This can be measured as a voltage $s_i(t)$ at receiver $i$ that is sampled and digitized at regular times $t$. Whereas a single receiver can measure the source brightness $I(\vec{d})$ in a specific direction $\vec{d}$, a pair of receivers $i,j$ separated by a baseline vector $\vec{B}_{ij}$ can be used as an interferometer to measure the difference in phase between the signals $s_i$ and $s_j$ due to the time delay $\tau_{ij}=\vec{B}_{ij}\cdot\vec{d}/c$ between the received signals as illustrated in Figure~\ref{interferometer}.

\begin{figure}[ht]
\centering
\includegraphics[scale=1.0]{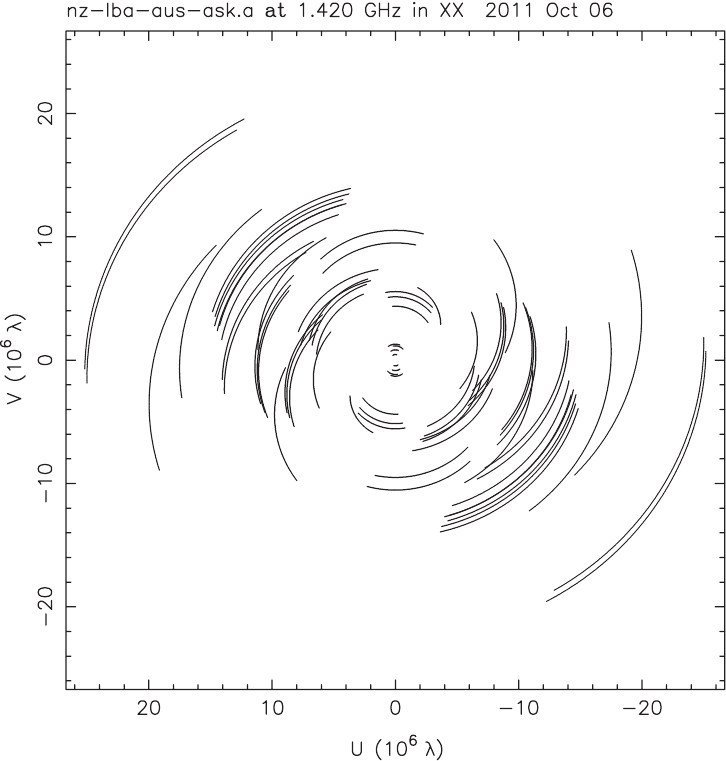}
\caption{A uv-plot shown for the VLBI array in Australia and New Zealand, including Australian Long Baseline Array (five radio telescopes), Warkworth (New Zealand), ASKAP (Western Australia) and AuScope antennas in Katherine and Yarragadee. The uv-plot results in $36\times 2$ baseline tracks for a $4$ hour observation, $21$-cm wavelength and common source declination of $-70^{\circ}$ \cite{sweston}.} 
\label{EarthRotationSynthesis}
\end{figure}

The time delay $\tau_{ij}$ can be roughly approximated by the geometry of the antennae relative to the source direction (provided by an \emph{Ephemerides service}), and more precisely determined by the resulting interference pattern in the cross-correlation between the signals:
\begin{displaymath}
  (s_i\star s_j)(\tau) = \int_{-\infty}^\infty \overline{s_i(t)}s_j(t+\tau)\,dt.
\end{displaymath}
The value $V_{ij}=(s_i\star s_j)(\tau_{ij})$ is termed a \emph{visibility} and gives a source brightness measurement at a $(u,v)$ point in the Fourier domain determined by the baseline $\vec{B}_{ij}$ for that pair of antennae. An array of $n$ antennae has $\frac{n(n-1)}{2}$ baselines (one per pair of antennae) and so $\frac{n(n-1)}{2}$ visibilities can be obtained. However, if readings are taken over an interval of hours then each baseline changes over time due to the rotation of the Earth. Hence, over time the baselines sweep out elliptical arcs in the Fourier plane, illustrated in Figure~\ref{EarthRotationSynthesis}.

The Fourier domain coverage of an array is the combination of the $(u,v)$ tracks from all baselines provided by the array. It shows where the array samples the Fourier transform of the source image. For high quality imaging, it is desirable to have the best possible coverage of the Fourier domain, which is effectively the telescope aperture. A perfect source brightness distribution (the image of the radio source) could be obtained simply by taking the inverse Fourier transform if all $(u,v)$ points in the Fourier domain were able to be measured, but this is never the case. A deconvolution process such as \emph{Clean} or \emph{Maximum Entropy} used in any modern interferometer imaging can be thought of as a scheme for interpolating or extrapolating from the measured $(u,v)$ points to all other points in the $(u,v)$ plane \cite{walker}.

\begin{figure}[ht]
\centering
\includegraphics[scale=1.0]{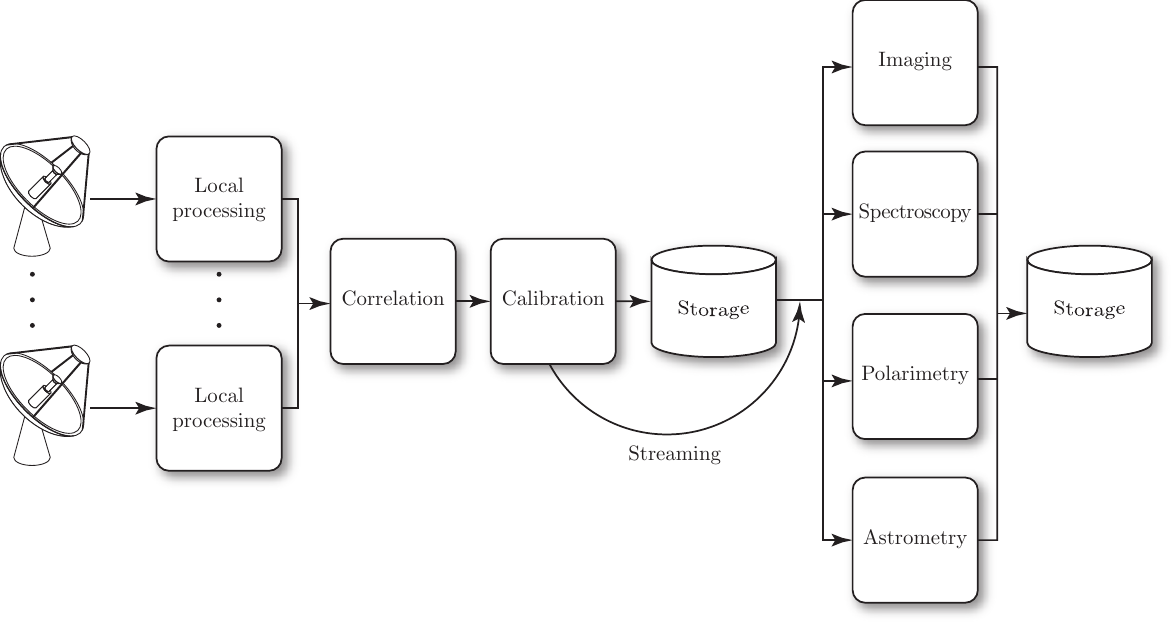}
\caption{Radio astronomy array pipeline}
\label{pipeline}
\end{figure}

Measuring the signals from all $n$ antennae over a period of hours results in an enormous dataset for a large array and its processing is a very compute intensive problem. Figure~\ref{pipeline} shows the operations that are performed in an array on the digitized signals in a simplified pipeline from raw data through to analyzed data products. Computational power required for these operations can be very significant, particularly for the Correlation operation which calculates the visibility $V_{ij}$ for each of the $\frac{n(n-1)}{2}$ baselines at each time $t$ via cross-correlations. It also performs an autocorrelation of each signal as discussed in Section \ref{sec:AutocorrelationPipeline}, which together with the visibilities forms the \emph{datacube} for the array at time $t$.

It is estimated that LOFAR with its 36 antennae stations can produce over 100 TB/day \cite{schaaf}. For the SKA which will eventually have about 3000 antennae dishes, the data will increase by at least 5 orders of magnitude \cite{varbanescu}. Such a huge amount of data places very high processing demands and requires a special approach to the overall organization of how data are processed and stored. It is only feasible to store the digitized raw signals required for calculating data cubes for small arrays and is limited to measurements taken over short time periods; in all other cases the data storage requirements are too large to be practical.

In the next section we introduce the stream computing paradigm and how IBM's InfoSphere Streams data management middleware utilizes this paradigm. In Section \ref{sec:InfoSphereStreams} we describe how InfoSphere Streams can be applied to the operational facets of large radio astronomy telescope arrays to handle the enormous data volumes and compute intensive operations. In Section \ref{sec:AutocorrelationPipeline} we consider an actual InfoSphere Streams application that performs streaming autocorrelations of actual radio astronomical observations.

\section{IBM InfoSphere Streams and the Stream-Computing Paradigm} \label{sec:InfoSphereStreams}

\begin{figure}[ht]
\centering
\includegraphics[scale=1.0]{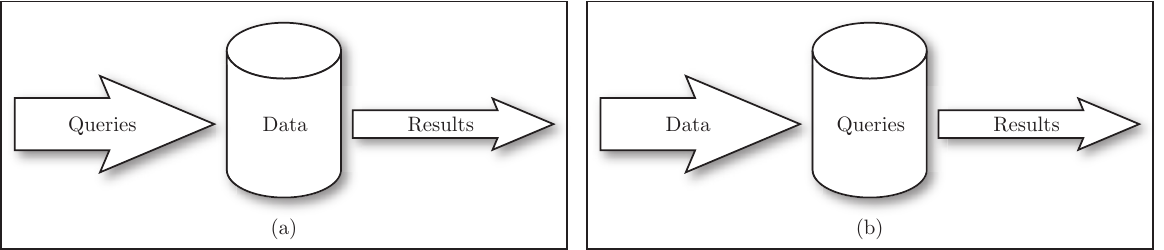}
%
%
\caption{(a) Traditional computing techniques versus (b) stream-computing paradigm \cite{streamssysinf}}
\label{streamcomp}       
\end{figure}

With the vast expansion of data volumes generated in the current information age, there has been a paradigm shift in data management toward the processing of streaming data. Stream computing differs from traditional computing in that real-time data is processed on the fly by  relatively static queries that continuously execute during the lifetime of an application, instead of the data being considered relatively static and all queries being short lived. This is illustrated in Figure~\ref{streamcomp}.
\newline\indent In May 2010 IBM released \emph{InfoSphere Streams} or \emph{Streams}. Streams is the result of several year's research conducted by the exploratory stream processing systems group at the IBM T.J. Watson  Center. It is a data stream management system middleware designed to ingest, filter, analyze and correlate enormous amounts of data streaming from any number of data sources. Streams is designed to facilitate a rapid response to changing environments leveraging the stream computing paradigm. It has the following objectives \cite{streamsdatamotion}:
\begin{itemize}
\item Scale using a variety of hardware architectures as demand for processing power changes.
\item Provide a platform for handling data streams that is responsive to dynamic user requirements, changing data, and system resource availability.
\item Incremental tasking for changing data schemes and types.
\item Secure transmission of data streams at all system levels, along with comprehensive auditing of the execution environment.
\end{itemize}

In the remainder of this paper, we focus on Streams version 1.2\footnote{The current version is 2.0 with similar philosophy  but with changes in the programming language} which was the platform used in our exploration.
\subsection{InfoSphere Streams terminology and concepts}

\begin{figure}[ht]
\centering
\includegraphics[scale=1.0]{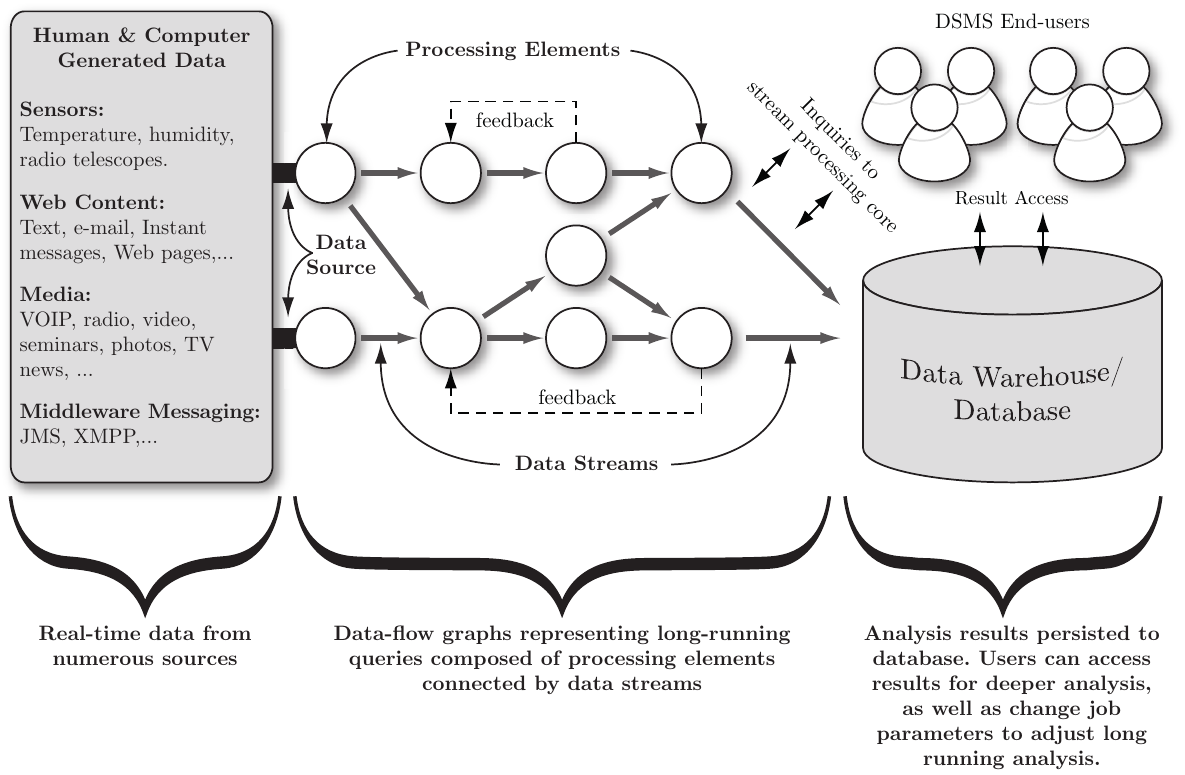}
%
%
\caption{Stream processing core executes numerous long running queries referred to as jobs, which are represented by data-flow graphs \cite{streamssysinf}}
\label{streamoverview}       
\end{figure}

Streams is designed to be highly scalable, so it can be deployed on a single node or on thousands of computing nodes that may have various hardware architectures. The \emph{stream processing core} distributed runtime environment executes numerous long running queries, which Streams refers to as \emph{jobs} \cite{streamssysinf}. A job can be represented by a \emph{data-flow graph}. Each vertex in the graph represents a \emph{processing element} that transforms the data, and each connecting edge is a data stream, as illustrated in Figure~\ref{streamoverview}.
\newline\indent Stream processing elements provide running statistics on their operation. These statistics are utilized by the stream processing core to dynamically optimize job performance by distributing the load and allocating suitable resources for executing each job \cite{rmfordsms}. Note that a processing element that maps to an underlying computing resource may be changed dynamically by the stream processing core according to load distribution.\newline\indent The following is a brief description of the stream processing core's main architectural components \cite{benchspc}, which is also illustrated in Figure~\ref{spcruntime}:
\begin{figure}[ht]
\centering
\includegraphics[scale=1.0]{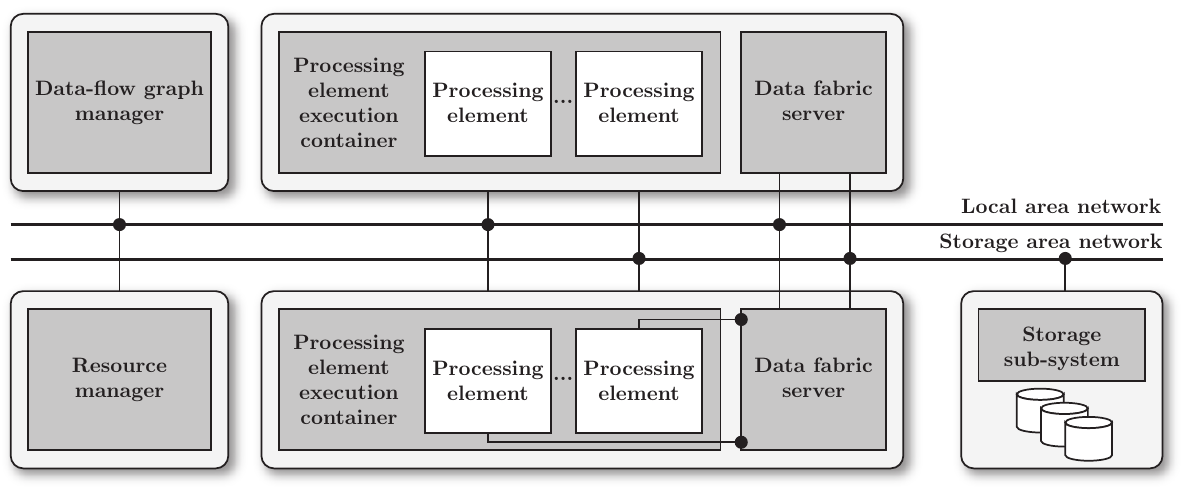}
%
%
\caption{Stream processing core distributed runtime environment \cite{spade}}
\label{spcruntime}       
\end{figure}
\begin{itemize}
\item \textbf{Dataflow Graph Manager}\newline
The \emph{dataflow graph manager} is responsible for the data stream links between the processing elements. Its primary function is to manage the specifications of the input and output ports.
\item \textbf{Data Fabric}\newline
The \emph{data fabric} provides the distributed facet of the stream processing. It is made up of a set of daemons that run on each available computing node. The data fabric uses the data stream link specification information from the data-flow graph manager to establish connections between the processing elements and the underlying available computing nodes to transport stream data objects from producer elements to consumer elements.
\item \textbf{Processing Element Execution Container}\newline
The \emph{processing element execution container} provides the runtime environment and access to the Streams middleware. Furthermore, it also acts as a security fence preventing applications that are running on the processing elements from corrupting the middleware as well as each other.
\item \textbf{Resource Manager}\newline
The \emph{resource manager} facilitates system analytics by collecting runtime information from the data fabric daemons and the processing element execution containers. The analytics information is used to optimize the operation of the entire system.
\end{itemize}

There are three different ways that users and developers can utilize Streams to process streaming data \cite{systems}:
\begin{itemize}
\item \textbf{Inquiry Services Planner}\newline
This level is designed for users with little or no programming experience. The inquiry services planner gives user access to a collection of predefined processing elements that generate underlying data-flow graphs (behind the scenes the planner generates SPADE applications).
\item \textbf{Stream Processing Application Declarative Engine (SPADE)}\newline
SPADE is an intermediate declarative language that enables the construction of data-flow graphs from predefined and custom stream operators.
\item \textbf{Streams API and the Eclipse Plug-in}\newline
This is designed for experienced developers who use programming languages such as C++ or Java to implement stream applications that run on the processing elements using the Streams API. Development can be facilitated by using a plug-in available for Eclipse.
\end{itemize}

\subsection{Data streaming applications with SPADE}

\begin{figure}[ht]
\centering
\includegraphics[scale=1.0]{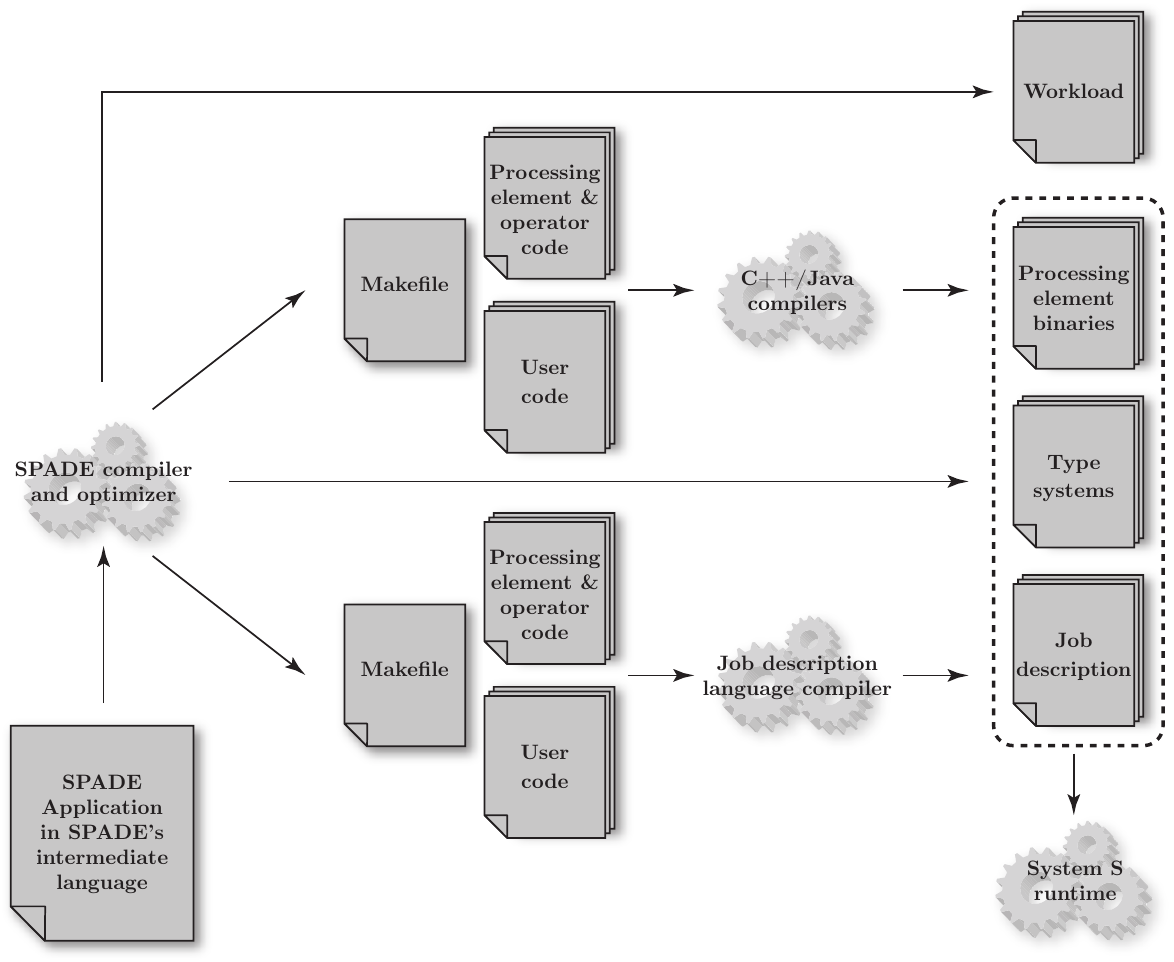}
%
%
\caption{SPADE's code generation framework \cite{spade}}
\label{spadecodegen}       
\end{figure}

Constructing a distributed stream processing application can be a complex process. The following considerations need to be made:
\begin{itemize}
\item What data stream transform operations must be developed. Transform operations are the building blocks that are combined together to ingest, process, analyze and produce the desired output data stream.
\item How the data stream transform operations can be mapped efficiently to distributed computing resources.
\item The interconnections, network protocols, scheduling and synchronization of operations between the available computing resources.
\end{itemize}
SPADE is designed to deal with these considerations so that programmers can focus on the design of a distributed stream processing application. Using SPADE they can avoid having to develop transform operators as well as face deployment issues that vary depending on the availability of computing resources, network infrastructure and specific technologies \cite{spade}. SPADE fulfills its design objectives by collaborating with the stream processing core to provide a dynamic runtime code generation framework capable of achieving scalability and performance through automatic deployment and optimization. This is illustrated in Figure~\ref{spadecodegen}.

The source code of a SPADE application is structured into five main parts:
\begin{itemize}
\item \textbf{Application Meta Information}\newline
This part contains the application name and optionally the debug/trace level.
\item \textbf{Type Definitions}\newline
In this part the name-spaces and aliases used by the application are declared.
\item \textbf{External Libraries}\newline
References to external libraries and header files that contain custom user defined operations are declared in this part. This part is optional.
\item \textbf{Node Pools}\newline
In this part pools of computing nodes can be optionally declared. This part is optional since the SPADE compiler can interact with the resource manager to discover available computing node resources.
\item \textbf{Program Body}\newline
This is the part where the actual SPADE application is written. In SPADE, streams are considered first class objects where the order of execution is fully characterized by the resulting data streams.
\end{itemize}

The SPADE language offers the following relational stream operators, used to construct long-running queries:
\begin{itemize}
\item \textbf{Functor}\newline
A \emph{functor} operator is used to carry out tuple level operations such as filtering, projection, mapping, attribute creation and transformation. A Functor can also access tuples that have appeared earlier in the input stream.
\item \textbf{Aggregate}\newline
An \emph{aggregate} operator facilitates grouping of input stream tuples. Tuples can be grouped in a variety of ways.
\item \textbf{Join}\newline
The \emph{join} operator is used for combining two streams in a variety of ways.
\item \textbf{Sort}\newline
The \emph{sort} operator is used to order tuples.
\item \textbf{Barrier}\newline
The \emph{barrier} operator is used for stream synchronization. It accepts tuples from multiple input streams and only starts to output tuples when it has received a tuple from each input stream.
\item \textbf{Punctor}\newline
A \emph{punctor} operator is somewhat similar to a functor operator. The difference between the two is that a punctor operator performs tuple level operations on the current tuple or tuples that have appeared earlier based on punctuations inserted in the data stream.
\item \textbf{Split}\newline
A \emph{split} operator is used to pass input stream tuples to multiple output streams based on specified user conditions.
\item \textbf{Delay}\newline
The \emph{delay} operator allows a time interval to be specified for delaying a data stream.
\item \textbf{Edge Adapters}\newline
Edge adapters are stream operators that function on the boundaries of the SPADE application. They allow a SPADE application to obtain and provide streamed data to applications and entities that are external to the system. There are two types of edge adapter operators:
\begin{itemize}
\item
A \emph{source} operator is used to create an incoming data stream of tuples from external data sources.
\item
A \emph{sink} operator is used to convert tuples to a format suitable for applications and entities that are external to the system, such as a file system, database, or external application.
\end{itemize}
\item \textbf{User Defined Operators (UDOPs)}\newline
SPADE allows external libraries to be utilized within the SPADE application. Functionality of existing operators can also be extended using UDOPs. UDOPs are developed in C++ or Java using the Streams Eclipse plug-in. UDOPs can be used to port legacy code from other data management platforms into the Streams platform. Furthermore, UDOPs can be used to wrap external libraries from other systems so they can be interfaced with the Streams platform.
\item \textbf{User Defined Built-in Operators (UBOPs)}\newline
Although UBOPs  allow users to define customized operators they are restricted to the scope of the SPADE application that declares them. On the other hand once defined UBOPs become part of the SPADE language and essentially available for use with any SPADE application.
\end{itemize}

SPADE also offers advanced features to extend its capabilities and provide a richer platform for data stream application developers.
\begin{itemize}
\item \textbf{Matrices, Lists and Vectorized Operations}\newline
Lists and matrices plus the capability to carry out operations on them is a core fundamental feature in many applications such as signal processing, computer graphics, data mining and pattern classification. SPADE offers native support for list and matrix data types as well as vectorized operations which operate on them. Lists or matrices can be created either from external sources via the source operator, functor or punctor operators can be used to create lists or matrices  from incoming tuples, or the aggregate operator can create lists or matrices from multiple tuple streams. Many of the SPADE built-in functions are capable of handling matrix, list and scalar type attributes.
\item \textbf{Flexible Windowing Schemes}\newline
SPADE supports general windowing mechanisms such as sliding and tumbling windows. SPADE takes these mechanisms further by allowing more sophisticated windowing mechanisms. As an example, an operator can accumulate tuples in a window to hold prior to processing. When a punctuation symbol is received, a processing operation is triggered on tuples currently contained in the window, such as averaging or summing the tuples, and then the window is made to tumble or slide.
\item \textbf{Per-group Aggregates and Joins}\newline
Per-group aggregates and joins are designed to cut the number of computations required for operating on a large number of tuple groups. SPADE has the ability to define distinct groupings within a window, so that when a trigger is received an aggregate or join operation can be applied to the entire window or distinct groups within the current window.
\end{itemize}

\subsection{Deploying SPADE applications and performance optimization}

Discovering the exact optimal mapping (deployment) of a parallelized computer program to loosely coupled (gridded) computing resources is an NP-hard problem \cite{jesper03}. However, heuristics techniques can be used as a practical means for determining an acceptable approximation to an optimal mapping \cite{culler93}. These heuristics can be improved over time by collecting running statistics that monitor the utilization and performance of computing and network resources.

A SPADE application is a parallelized computer program since it consists of many operators working in parallel towards achieving a common task. InfoSphere Streams approaches optimization of mapping a SPADE program to its underlying computing and network resources in two stages. First, how  operators are logically combined (\emph{fused}) into processing elements, and second how processing elements are assigned to physical computing nodes \cite{gedik09}.

Info Sphere Streams uses a profiling framework that repeatedly maps processing elements to physical nodes, collects statics and makes necessary remapping adjustments. At the same time the \emph{fusion optimizer} uses the collected statics along with a greedy algorithmic technique to fuse operators into single processing elements.

\section{Utilizing InfoSphere Streams to Address Large Antennae Array Software Architecture}
\label{sec:LargeAntennae}

One of the characteristics of radio astronomy is that it often involves very large volumes of data, particularly when an array of radio telescopes is used for radio interferometry to obtain greater angular resolution of a celestial object. It also has involved many ad-hoc techniques for processing and managing the data.

\begin{figure}[ht]
\centering
\includegraphics[scale=1.0]{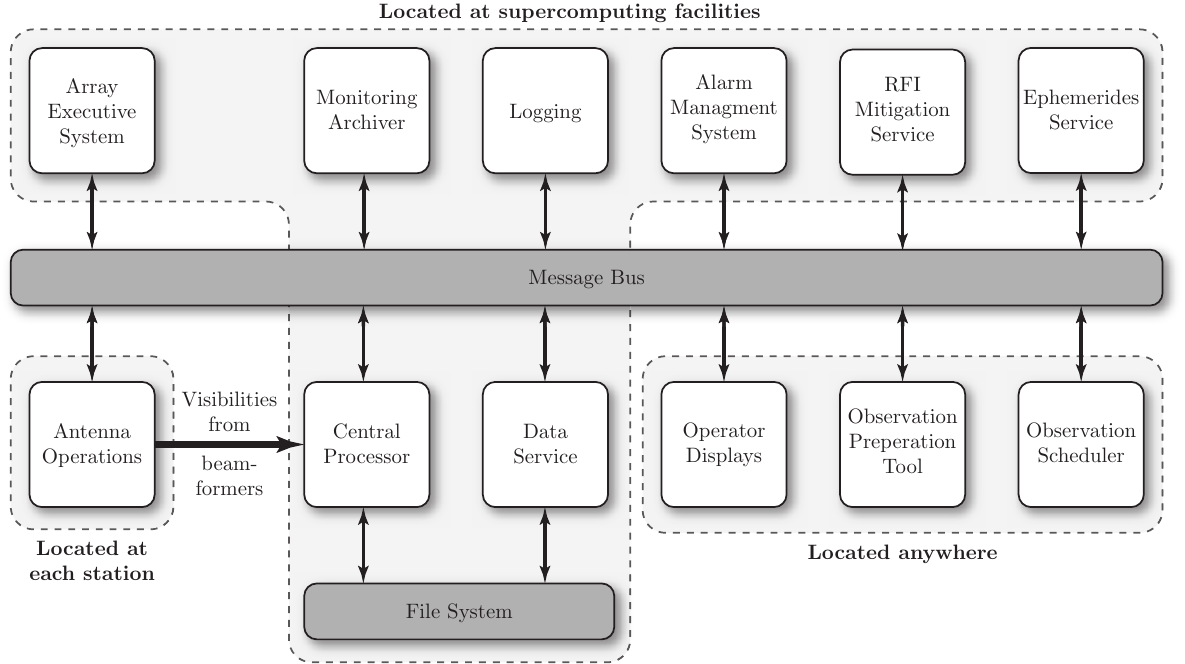}
\caption{ASKAP top level architectural component view adapted from \cite{guzman10}}
\label{fig:ASKAParchitecture}
\end{figure}

For instance, the Australian Square Kilometre Array Pathfinder (ASKAP) is a CSIRO-led radio telescope array currently under development at the Murchison radio astronomy observatory. It will consist of 36 antennae, each with a phased array feed that supplies 1.9 Tbps of data and requires 27 Tflops processing to extract a beam visibility. Correlating the resulting 0.6 Tbps data from each antenna is estimated to require 340 Tflops and provide 8 Gbps results for further analysis. ASKAP will have the following the architectural components as given in Fig.~\ref{fig:ASKAParchitecture}, which illustrates components required to control and manage the data pipeline in a radio telescope array \cite{guzman10}:
\begin{description}
   \item[\emph{Antenna Operations:}]
   
   includes positioning an antenna and setting data acquisition parameters such as sampling rate, bit resolution and filter bank configuration.

   \item[\emph{Central Processor:}] 
   correlates the beam visibility data and performs further analysis such
   as image synthesis or spectral line work.

   \item[\emph{Array Executive System:}]
   responsible for coordinating an observation by the array.

   \item[\emph{Monitoring Archiver:}]
   archives monitoring data generated by the system.

   \item[\emph{Logging:}]
   responsible for logging messages generated by the system.

   \item[\emph{Data Service:}]
   responsible for managing the database storage.

   \item[\emph{Alarm Management System:}]
   manages alarm conditions such as failures.

   \item[\emph{RFI Mitigation Service:}]
   identifies potential sources of radio frequency interference in the received signals.

   \item[\emph{Ephemerides Service:}]
   calculates the positions of celestial objects.

   \item[\emph{Operator Display:}]
   a user interface for system control and monitoring.

   \item[\emph{Observation Preparation Tool:}]
   facilitates the setup and pre-planning of observations.

   \item[\emph{Observation Scheduler:}]
   generates schedules for the execution of observations by the Executive System.
\end{description}

Although initially InfoSphere Streams has primarily been applied to the analysis of financial markets, the healthcare sector, manufacturing, and traffic management, it is also suited for managing a radio astronomy pipeline. In particular, the ASKAP architecture in Fig.~\ref{fig:ASKAParchitecture} provides a high level abstracted view of how that data pipeline will look like within the ASKAP system. That data pipeline could be instantiated with minimal design effort into streams operators graph, making the transformation of the data in the pipeline transparent. It also allows for the mapping of stream operators to processing elements to be dynamically reconfigurable, important for system scalability, optimizations, and fault tolerance. In fact most of the described architectural components would benefit from these features.

One straightforward scenario for a streams software design mapping the ASKAP architecture is to implement two core main streams instances. The first instance runs at the front-end close to the antennae and is responsible for data conditioning, RFI mitigation, and visibility production. The second instance runs at the central processing unit, and is responsible, among others tasks, for generating images from the visibilities. These two Streams applications could communicate through the Streams middleware services and implement a fast, real-time processing scheme for managing the data from its acquisition all the way through to the analyzed data products.

Other streams jobs could run on the central processing unit. One such job could be responsible for logging, archiving and storage using sink operators. Another could communicate with the scheduling system and send control signals for coordination.

\subsection{Data provenance and management capabilities}

Radio astronomical artifacts such as visibilities can be characterized by the observation schedule, station configuration, along with the recording and processing performed on the sampled receiver voltages. This metadata describes how an artifact has resulted from an observation and so provides important provenance information. The common provenance standard for VLBI  is the VLBI EXperiment (VEX) format. A VEX file provides a complete description of a VLBI experiment from the scheduling and data capture through to the correlations that result in a data cube. It is designed to be independent of any data acquisition and correlator hardware and to accommodate new equipment, recording and correlation modes \cite{vexvlbi}. Every VEX file starts with a line identifying the file type and VEX version, and is followed by a number of separate blocks, currently classified as either:

\begin{itemize}
\item Primitive blocks which define low-level station, source, and recording parameters, such as antenna configuration and clock synchronization.
\item Global block which specifies general experiment parameters.
\item Station and mode blocks which define keywords that combined with the global parameters provide a detailed configuration for an observation at a station.
\item Sched block which specifies an ordered list of observations to perform.
\end{itemize}
For a steady-state radio source the VEX file allows a VLBI experiment to be reproduced, which can be valuable in large arrays where it is impractical to store the huge volumes of raw data. For transient or micro-lensing observations \cite{McLaughlin,Refsdal} the VEX file provides a basic audit trail for verifying the origins of the experimental results.
 
However, data provenance practices are less standard for stages after correlation processing. Correlated data can be analyzed in a variety of way, such as for image synthesis or to obtain the power spectral density, and there are not yet standard formats for defining how the resulting data artifacts are produced. The development of a set of provenance standards will be essential for the SKA to ensure the origins of the large number of artifacts produced and for automating their generation.
 
A radio astronomy system such as ASKAP should be able to provide the following characteristics for managing data:
\begin{itemize}
\item Adequate end-to-end throughput not hindered by latency due to the processing elements.
\item Intermediate storage capabilities (persistence) to be able to store data summaries in a storage of choice, such as a database or file.
\item High availability: the system should be able to work reliably and if failure occurs, data recovery services should be available to avoid important and critical loss of data.
\end{itemize}

In Streams,  adequate throughput  can be achieved  by a proper architecting of the operators' graph (the graph illustrating how  processing elements are linked together to form an application)  and by an optimal assignment of processing elements to nodes.  This process can only be done on an application-by-application basis. IBM InfoSphere Streams facilitates the design and optimization of such a graph. Optimizations can be done at compile time, such as those related to the placement of processing elements into nodes or the fusion of processing elements into operators, or at run-time, such as when some nodes get overloaded.  Compile-time optimization is efficient when workload and underlying resources are static. Offline profiling of system characteristics prior to deployment is also available, in which case a first pass (prior to deployment) can provide statistics on the data throughput and a second pass (on-deployment) uses those statistics to optimize operators placement and determine when fusion of processing elements should occur.

Data storage can be achieved by use of sink operators  capable of storing data to a file, database, or url port. In addition, the software supports user-developed sink operators, useful for custom-based storage needed when sending data to specialized storage recipients.

High availability can be achieved at the middleware level, the application level, and the operator level. At the middleware level, various services are provided to restart a job (potentially on a different node or hardware), replicate name servers across multiple nodes, and  monitor activities by writing log files to transactional storage recipients. At the application-level (data processing level), Streams provides checkpointing and automatic restart of processing elements in case of failure. It has tools to provide partial fault tolerance when data loss is a critical issue by means of state persistence capability (the capability to save the state of an operator and restore it).

\subsection{Some applications of Streams in radio astronomy}
There are several examples where Streams has been utilized in the area of radio astronomy and space science.

A space weather monitoring system was developed through joint work by LOIS and IBM Research  \cite{lars09}. It is known that the high-rate, large-volume of near-Earth space data generated by various satellites (such as those of the European Space Agency)  is a serious challenge for standard techniques for space weather data monitoring and forecasting. In particular, mining these data in a store-and-process system is not amenable. Streams software was used in \cite{lars09} to develop a real-time streaming application that  measures the intensity, polarization and direction of arrival for signals in the 10 kHz and 100 MHz frequency bands, and on-the-fly generated signal summaries that could be used for space weather forecasting and prevision.

A streaming version of the convolution resampling algorithm was developed by IBM and CSIRO \cite{biem10} as a prototype imaging application in the Central Processor of ASKAP as described earlier. The version of the algorithm implemented was the w-projection algorithm, which included a CPU intensive gridding step (the process of mapping visibility coordinates into a power of 2 grid). That study showed the flexibility of the streaming software by describing various implementations of streaming scenarios resulting in significant improvements in gridding time.

The next section describes a stream-based autocorrelation approach developed using Streams for data received from a single radio telescope.

\section{Implementing a Stream-Centric Autocorrelation Data Pipeline \& Utilizing Hardware Accelerators}
\label{sec:AutocorrelationPipeline}

Cross correlation is a fundamental tool in radio astronomy since it helps with identifying repeating patterns obscured by the predominant random noise content of extraterrestrial signals. As these signals are mostly composed of random noise they can be characterized as stationary stochastic processes where the mean and variance do not change over time.

An autocorrelation is the cross correlation of a signal with itself. Autocorrelation is mainly used for single antenna applications and calibration of individual arrayed antennae. Implementing autocorrelation requires less effort since only one signal is considered and therefore no time delays are required. As a consequence of its relative simplicity implementing an autocorrelation pipeline is a logical starting point for constructing a basic cross correlation pipeline for radio astronomy signal processing.

\subsection{Autocorrelation and the power spectral density in radio astronomy}
\label{subsec:AutocorrelationPSD}

If voltage samples $s(t)$ are obtained from an antenna the energy spectral density $E(f)$ of the incident electromagnetic waves can be determined. The energy spectral density is the energy carried by the incident waves per unit frequency $f$, which is given by the Fourier transform:
\begin{displaymath}
E(f) = \left|\int_{-\infty}^\infty s(t)e^{-2\pi i f t}\,dt\right|^2.
\end{displaymath}
However, $s(t)$ is a stationary signal and is not square integrable so its Fourier transform does not exist. Instead the \emph{Wiener-Khinchin} theorem is applied to obtain the power spectral density (PSD) of the voltage signal from the autocorrelation function $r(\tau)$:
\begin{displaymath}
r(\tau)=<s(t)s(t-\tau)>.
\end{displaymath}
The Wiener-Khinchin theorem states that the PSD $P(f)$ of the signal $s(t)$ is the Fourier transform of the autocorrelation function $r(\tau)$:
\begin{displaymath}
P(f)=\int_{-\infty}^{\infty}r(\tau)e^{-2\pi i f t}\,dt.
\end{displaymath}
The PSD $P(f)$ is the power carried by the incident waves per unit frequency $f$.

\begin{figure}[ht]
\centering
\includegraphics[scale=1.0]{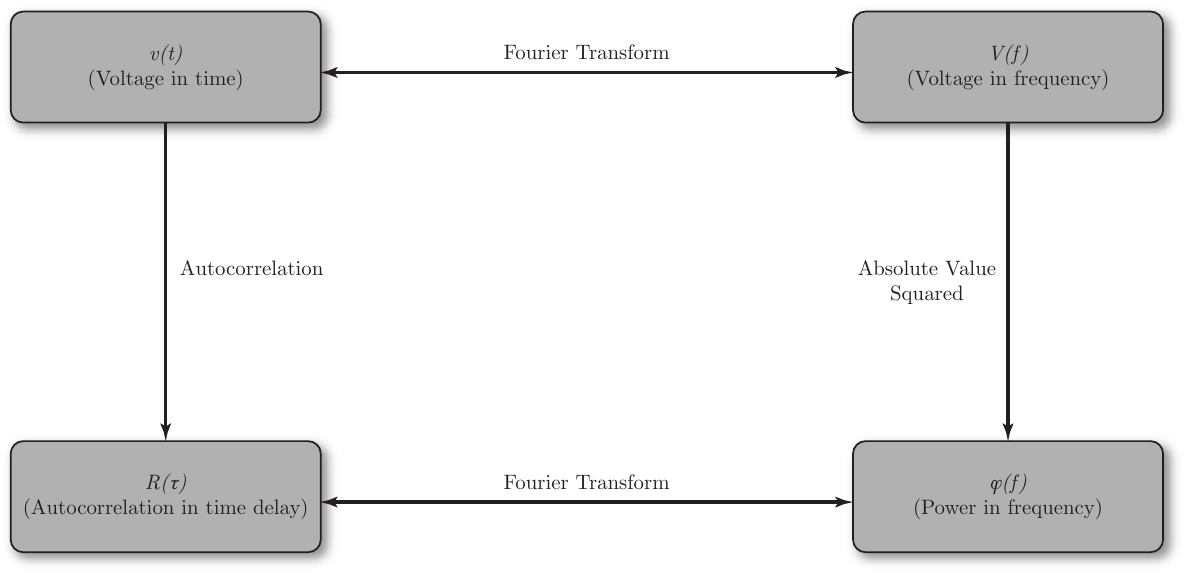}
%
%
\caption{Relation between voltage in time and frequency domains with the autocorrelation function and power spectral density \cite{toolsRA}}
\label{autocorrpsdrelation}       
\end{figure}

From Figure~\ref{autocorrpsdrelation} the PSD can be obtained by either performing an FX or XF correlation. An FX correlation is a Fourier transform followed by element-wise multiplication. An XF correlation is a cross multiplication followed by a Fourier transform. FX style correlation is preferred for software implementations since it involves fewer multiplications \cite{bunton05}.

\subsection{Implementing a PSD pipeline as a stream based application}
\label{subsec:ImplementingPSDStream}

\begin{figure}[ht]
\centering
\includegraphics[scale=1.0]{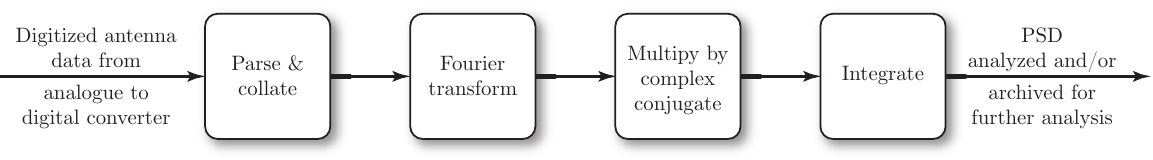}
%
%
\caption{FX PSD pipeline data flow}
\label{fxpsdflow}       
\end{figure}

Analogue voltage signals on the antenna receiver are sampled and digitized by an analogue to digital converter. The digitized real value data (2-16 bit digitization) are then streamed in real-time into an FX style pipeline to produce the power spectral density (PSD) of the signal. The FX PSD pipeline illustrated in Figure~\ref{fxpsdflow} is comprised of the following steps:

\begin{itemize}
\item Collect digitized signal data into chunks whose size is determined by the amount of data optimally processed together in the pipeline.
\item \emph{Channelize} each chunk to obtain frequency domain data by applying a Fast Fourier Transform to obtain single-precision float complex value data chunks.
\item Obtain the autocorrelation of the data in the frequency domain by multiplying each complex value in a chunk by its complex conjugate.
\item Integrate and average the data chunks over time to obtain a best PSD estimate of the signal.
\end{itemize}

The final stage plays an important role in improving the signal to noise ratio, hopefully allowing buried coherent signals of interest to emerge from the predominantly random noise polluted signal.

The entire pipeline can be viewed as a parallel program where each stage of the pipeline is an independent task. As data flow through the pipeline each stage operates on the data concurrently. Since the auto-correlation pipeline can be decomposed into independent stages the pipeline can be easily defined as a Streams application with appropriate SPADE operators.

Compute intensive tasks can be delegated to specialist hardware accelerators such as GPUs, PowerCell CPUs or FPGAs. Delegating tasks to various computing architectures demonstrates Stream's capabilities to construct and deploy parallel programs to heterogeneous computing clusters. For the PSD pipeline the most compute intensive task is the Fourier transform, and will be assigned to a hardware accelerator for processing.

The SPADE application makes use of \emph{virtual streams} for predefining the various tuple data structures, known as a tuple's \emph{schema}, that are utilized by its underlying stream operators. Virtual streams contribute towards ease of programming as well as understandability of the tuple structure flowing between stream operators. The PSD SPADE application declares the following virtual streams:

\begin{description}
\item[\texttt{RawData(data:ShortList)}]
defines the schema used for creating tuples arising from ingesting and parsing real value integer radio astronomy antenna data. The bit resolution used to digitize the antenna analogue signal may vary from as low as 2 bits up to 16 bits according to the recording system and type of observation. Essentially, using a \texttt{ShortList} data type satisfies the bit-level representation requirements for most radio astronomy recording formats.

\item[\texttt{RawDataChunk(acceleratorID:Integer, schemaFor(RawData))}]
defines the structure for a data chunk designated to a specific accelerator server for channelization.

\item[\texttt{ChannelData(real:FloatList, imag:FloatList)}]
defines the structure for channelized data chunks that have been channelized by an accelerator server.

\item[\texttt{PowerSpectrumData(psd:FloatList)}]
defines the structure for tuples containing PSD data. This virtual stream's schema is used by several operators for producing the different integration stages of the PSD.
\end{description}

Altogether the autocorrelation spectrometer SPADE application uses seven distinct stream operators. The number of actual operators depends on the number of accelerators utilized for channelization and integration stages. Figure~\ref{psdspade} shows the data flow graph for the SPADE PSD application (see Appendix for SPADE code listing).

\begin{figure}[ht]
\centering
\includegraphics[scale=1.0]{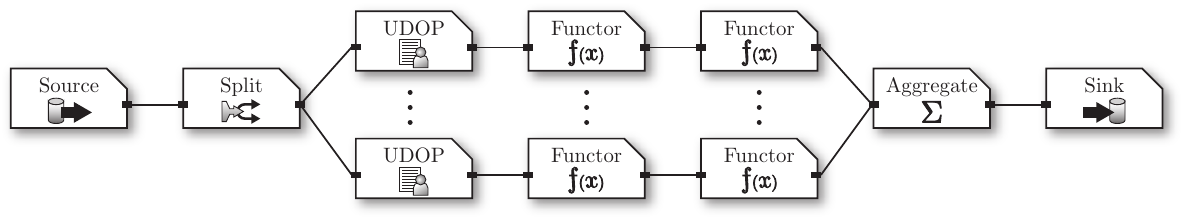}
%
%
\caption{FX PSD pipeline data flow graph and SPADE operators}
\label{psdspade}       
\end{figure}

The first stream operator in the application is a Source operator which ingests digitized unsigned integers from the signal. The data are parsed by this operator according to the format used to digitize and pack samples. The Source operator builds a tuple according to the \texttt{RawDataChunk} virtual stream schema, reading into the \texttt{ShortList} data tuple attribute. Each \texttt{RawDataChunk} tuple is assigned to an accelerator server by assigning a positive integer to the \texttt{acceleratorID} tuple attribute. The number of tokens contained in the data attribute is given by the size and number of FFTs to be performed by the channelization stage. A positive integer value between 0 and the total number of accelerators utilized is assigned to the \texttt{acceleratorID} attribute in a round robin fashion.

Tuples resulting from a Source operator are ingested by a Split operator. Essentially the Split operator is a multiplexed stream. Each sub-stream in the multiplex stream carries \texttt{RawDataChunk} tuples according to their respective identifier. Effectively the Split enables \texttt{RawDataChunk} tuples to be fanned out to several accelerators.

The Streams Processing Core may not be supported on a particular accelerator architecture, so one way around is to use a UDOP. To enable asynchronous communication with the accelerator server multi-threaded UDOPs (MTUDOP) are adopted by the SPADE application. An MTUDOP facilitates uncoupling the processes of receiving and transmitting tuples. For further versatility the UDOP uses configuration switches so that the same UDOP can be reused.

The MTUDOP uses three switches allowing the SPADE application to configure its operation with respect to which accelerator will be used for processing as well as the communication mode for incoming and outgoing data. During the initialization phase the MTUDOP extracts configuration information from the switch operators and establishes incoming and outgoing connections.

Once the connections have been made to a specific accelerator the MTUDOP runs two processes:
\begin{itemize}
\item The input tuple process ingests tuples transmitted by a specific Split operator sub-stream. The ingested tuples are converted to floats since software implementations for Fourier transforms require this. Following that some byte reordering may be necessary depending on the architecture of the accelerator. Once the type conversion and byte reordering are accomplished the data are sent to the accelerator for channelization.
\item The output tuple process receives data from the accelerator. Similarly to the previous process, received data may need byte reordering. The \texttt{ChannelData} schema is used to define the outgoing tuple structure. The channelized data chunk received from the accelerator arrives in interleaved complex number format, and so real and imaginary parts are separated into two \texttt{FloatList} data types. One \texttt{FloatList} represented by the tuple attribute \texttt{real} holds real values and the other \texttt{FloatList} represented by the tuple attribute \texttt{imag} holds imaginary values.
\end{itemize}

\texttt{ChannelData} produced by the MTUDOP are ingested by a Functor operator responsible for computing the instantaneous PSD values, multiplying each complex number by its complex conjugate to produce a real value. The resulting real values are defined by the \texttt{PowerSpectrumData} virtual stream schema.

Each \texttt{PowerSpectrumData} tuple arising from the first Functor operator contains data for several FFT problems, so the second Functor operator integrates those results. Effectively the second Functor integrates multiple FFT problems contained in a single data chunk. To accomplish integration within a data chunk the second Functor operator uses customized user-defined logic. Integrations within a data chunk are performed using a \emph{Slice} operation that helps with extracting the result of each FFT problem for summing and averaging. Summing and averaging produces a \texttt{FloatList} and so the same \texttt{PowerSpectrumData} schema is used to represent the resultant tuples.

\texttt{PowerSpectrumData} tuples from the second Functor operator are then integrated with an Aggregate operator. The aggregation count is specified by the SPADE application. Fundamentally the aggregation count is the required integration time. The longer the integration time, that is the higher the aggregation count, the better the signal to noise ratio.

Integrated power spectral density tuples produced by the Aggregate operator and defined by the \texttt{PowerSpectrumData} schema are ingested by a Sink operator. The Sink operator may either write the integrated PSD strips to disk or possibly stream them over the network for deeper analysis or visualization.

\subsection{Using accelerators (heterogeneous computing)}
\label{subsec:UsingAccelerators}

An accelerator is intended to provide specialized accelerated computing services to assist with handling compute intensive operations. The objective for utilizing accelerators is to enable real-time data management operations especially for areas that involve processing large amounts of data. An important consideration that must be made when using a particular accelerator hardware is the use of its unique performance primitives and libraries. Neglecting this consideration in many cases will lead to ineffective utilization of the accelerator's intensive computing capabilities.

The SPADE PSD application described previously is designed to function with any type of accelerator. At the time of implementation the PowerCell CPU accelerator was available. In this subsection we describe how the compute intensive Fourier transform was implemented on the PowerCell CPU using its unique performance primitives and libraries.

The Fourier transform service is provided by an implementation of the discrete Fast Fourier Transform (FFT) on a PowerCell QS22 Blade Server. The QS22 Blade Server comprises of two PowerCell CPUs. Each PowerCell CPU comprises of nine cores; one 64 bit duo-core PowerPC processor and eight 128 bit RISC processors.

\begin{figure}[ht]
\centering
\includegraphics[scale=1.0]{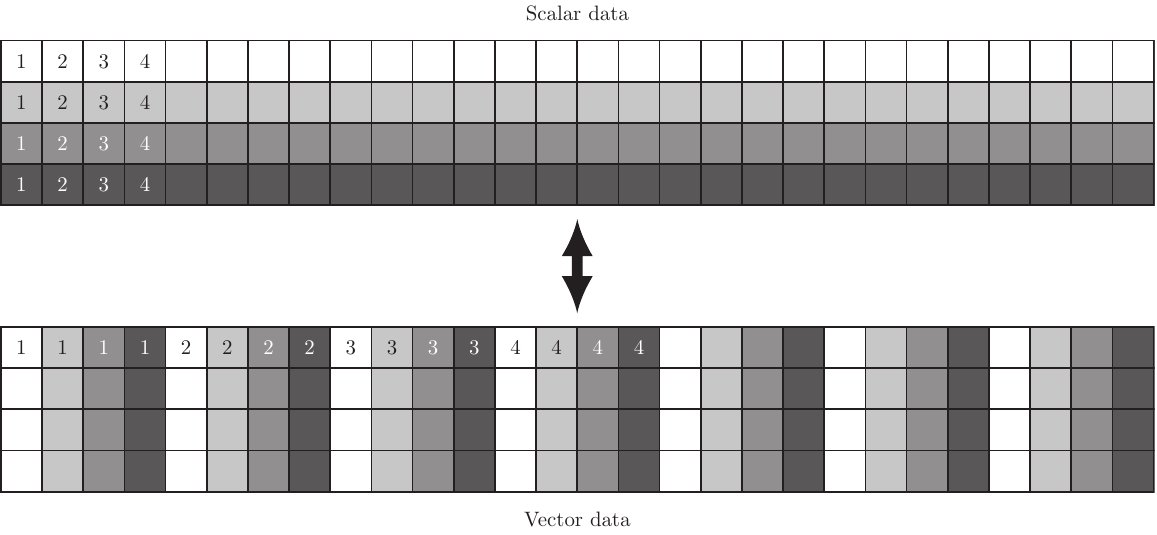}
%
%
\caption{Scalar versus vector data arrangement in a contiguous block of main memory \cite{cellredbook}}
\label{vecsynth}       
\end{figure}

The FFT server program is a multi-threaded application that executes the following four threads:
\begin{itemize}
\item Applications main thread; responsible for initializing the FFT memory buffers working area as well as starting the receiving, processing and sending threads. After initialization the main thread blocks until the application is terminated.
\item The receiving thread creates a server socket and listens for a single client connection. Once a sender client connects, the FFT Server receives data chunks containing multiple FFT blocks. Each data chunk is written to a specific buffer, which is then flagged to indicate that it is ready for processing.
\item The processing thread performs a real to complex FFT on memory buffers that have been flagged as fully received by the receiver thread. Consequently since the FFT is real to complex then ultimately the same amount of data received will be the same amount sent. The FFT is accomplished using all 16 SPU cores on a QS22, hence the reason why a single data chunk contains multiple FFT blocks. Once a single data chunk contained in a given buffer has been processed it is then flagged by the processor thread to indicate that the results are ready for sending.
\item The sending thread sends the contents of memory buffers that have been flagged as processed. Memory buffers that have been sent are flagged to indicate that the buffer can be reused for receiving.
\end{itemize}
The application makes use of a multi-buffering scheme for allowing the receiving, processing and sending threads to operate in an asynchronous fashion. The threads operate asynchronously as long as there are buffers available. Any contribution towards asynchronous operation between concurrent threads reduces blocking thereby contributing to a gain in overall performance. Nevertheless mitigating concurrency between thread access to each individual buffer is still required, and this is accomplished via a two-phase locking mechanism.
The processing thread makes use of the SDK for Multicore Acceleration FFT library to efficiently compute a large number of FFT problems in parallel. The FFT library achieves significant computational performance gains by exploiting the PowerCell CPU's vectorized SIMD capabilities utilizing two main approaches:
\begin{itemize}
\item \textbf{Striping across vector registers}\newline
The SPE's architecture is 128-bit hence its underlying Synergistic Processing Unit (SPU) registers can be considered as vector registers. In the case of this implementation the FFTs are performed using 32 bit (single-precision) floating point values. A single SPU register can therefore hold four individual 32-bit floats and operate on the entire vector of floats using a single instruction.
When performing FFT operations rather than loading four values from one FFT problem, four values are loaded from four problems. This technique is known as \emph{striping} multiple problems across a single register. Striping values from four problems across a register enables these problems to be accomplished in unison, and the code used for indexing and twiddle calculations can be reused \cite{cellredbook}.

The FFT algorithm makes extensive use of compute intensive trigonometric mathematical operations. Fortunately the SIMD Math Library contains vectorized versions of common mathematical operations. Utilizing vectorized math operations along with data stripping across vector registers dramatically reduces the frequency of their usage. However data striping values across vector registers is limited since it requires all the values from the FFT problems to be present in the SPU's LS, which has maximum capacity of only 256 kB. Hence data striping across vectors is limited to small point size FFT problems.
\item \textbf{Vector synthesis}\newline
To accommodate the memory bound limitations imposed by a SPU's LS, SPU \emph{shuffle} operations can rearrange a large set of FFT problems residing in main memory into vector form prior to DMA transfer to the SPUs. This rearrangement of scalar FFT problem values into vector form is known as \emph{vector synthesis}, and is illustrated in Figure~\ref{vecsynth}. Subsequently, once the SPUs complete FFT computations of the large set of problems, the vectorized values must be rearranged back into scalar form.
Naturally the rearrangement of a large data set of FFT problems into vectors then back to scalars does incur a computational expense. However since the large data set is prepared for vectorized trigonometric operations then the gains made in reducing the amount of computational intensive trigonometric operations greatly outweigh the costs incurred by vector synthesis operations \cite{cellredbook,cellFFT}.
\end{itemize}

To gain further significant performance speed-ups (of almost 10 times) on the Cell B.E. architecture, the implementation uses the huge translation lookaside buffer (TLB) file system. Huge TLB page files allocate large pages (16 MB per page) of contiguous memory. Utilizing huge page files reduces the TLB miss rate and consequently leads to a gain in performance. The data chunk size mentioned in Section \ref{subsec:ImplementingPSDStream} was set to 16 MB to fully utilize an entire huge TLB page.

\subsection{Testing the SPADE PSD application}
\label{subsec:Evaluation}

The SPADE PSD application was tested using network streamed data from the AUT University 12m radio telescope located at Warkworth in New Zealand. An IBM Blade Center holding x86 HS12 and dual-PowerCell QS22 blades was used to run the application. The HS12 blade was used to execute the SPADE PSD application. As the PowerCell FFT library is limited to a maximum size of 8192 points (for real to imaginary number FFT transforms) each huge TLB page could accommodate 512 FFT problems.

\begin{figure}[ht]
\centering
\includegraphics[scale=0.97]{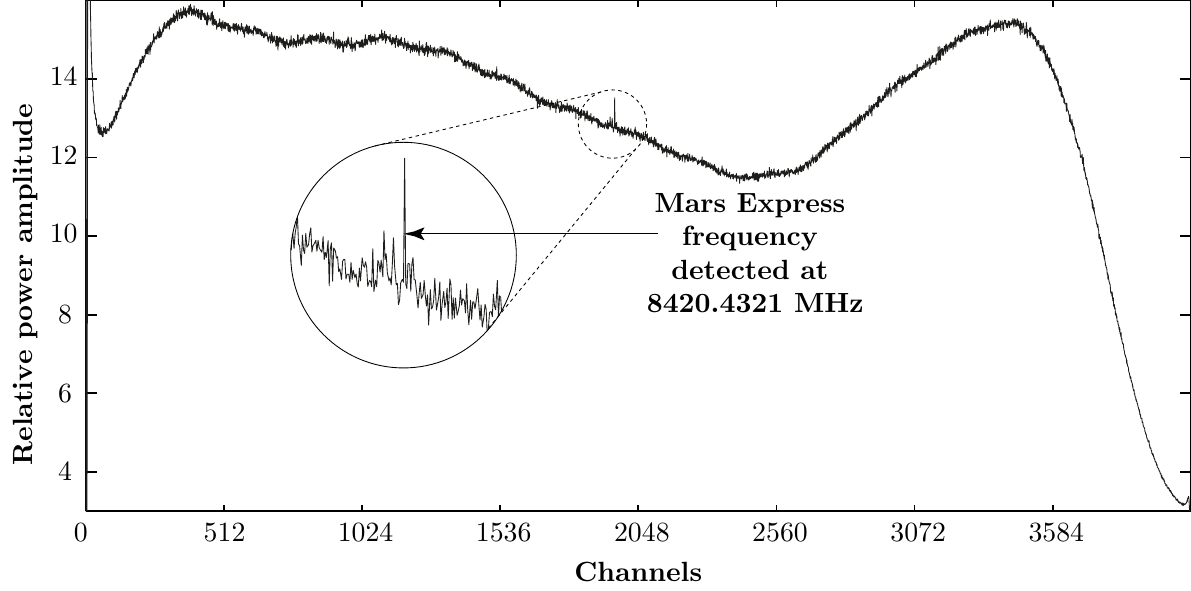}
%
%
\caption{Plot of the PSD results obtained from a 40 second observation of the ESA Mars Express spacecraft conducted by the AUT 12m radio telescope}
\label{spectrum}       
\end{figure}

Test data were sampled by the radio telescope from the European Space Agency Mars Express Orbiter at 60 MHz (32 MHz bandwidth due to Nyquist criterion) using 8-bit digitization. The resulting power spectral density is illustrated in Figure~\ref{spectrum}.

The overall shape of the distribution is dictated by the specific telescope and its receiving system. This shape was also determined independently using a hardware spectrum analyzer to verify the correctness of the PSD application. Of particular interest in the spectrum was the 8420.4321 MHz signal detected by the application which was being emitted by the orbiter at the time of the test.

On average the application took 8 ms to autocorrelate 512 FFT problems of size 8192 with a just single accelerator, a performance of 30 Gflops. Although greater performance could also be achieved by either using an FFT implementation tuned for the specific size or by utilizing more accelerators, the application could already handle 524 MHz sampling from a single antenna. However, careful attention must be paid to the networking layers in order to utilize the potential power of any accelerator. During initial testing the QS22 blade accelerators utilized TCP over full duplex Gigabit Ethernet links and both the average data chunk inter-arrival and inter-departure times were found to be 0.6 s, limiting the sampling rate to 6 MHz per link. This limitation can be removed by either utilizing multiple links in a round robin fashion from an x86 blade, utilizing an alternative protocol to TCP, or employing an alternative networking technology such as Infiniband.

\subsection{Performance and scalability}
\label{subsec:Performance}
The results from the previous subsection demonstrate that Streams shows good performance calculating the PSD using a mix of HS12 and QS22 blades with the PowerCell as the accelerator. In particular, the greatest performance limitation was determined to be inter-blade networking rather than anything associated with the Streams framework itself, despite Streams being hardware and network technology agnostic. Streams allows the underlying computer and network technologies to be changed for best suiting the computations required in a particular application. This enables it to leverage the performance of new hardware as it becomes available, while reducing the effort to reengineer software applications.


The scalability features of Streams are valuable for meeting growing computational demands. The SPADE PSD application can scale to handle greater sampling rates, higher frequency resolution via a larger FFT point size, or additional antennae. The SPC performs dynamic assignment of processing elements to physical nodes which enables the SPADE application to dynamically meet the demands of intensifying computations. This ability to dynamically redeploy a parallel program during runtime to physical nodes allows Streams to scale effectively.

\section{Conclusion}
\label{sec:Conclusion}

This application successfully demonstrated the viability of implementing a real-time  PSD entirely in software using InfoSphere Streams. The SPADE application showed good data throughput without being specifically tailored to a specific accelerator, and allowed dynamic reconfiguration to allow more accelerators to be utilized as necessary or alternative types of accelerators included. Due to the use of standard SPADE operations the management of the data through the pipeline was transparent and the application could be easily extended to provide further analysis or provenance features.

The operations of an FX style autocorrelating spectrometer pipeline are dominated by the algorithmic complexity $n\log_2(n)$ of the FFT operation. A single dual-PowerCell QS22 blade measured 30 GFlops for the FFT operation. In comparison measuring the performance of the FFT operation utilizing all four cores of a single x86 HS12 blade achieved almost 10 times less performance, and was measured at 3.5 GFlops. Essentially this shows that using various architectures for parallel computing by utilizing suitable accelerator hardware for specific compute intense operations can yield significant speed ups. In our case for an FX style autocorrelation pipeline a speed up of almost 10 times was achieved per QS22 blade.

IBM Info Streams proved it's flexibility to operate using various architectures in unison. Despite the I/O bound links Streams was capable of maximizing the link bandwidth as well as manage the data flow without information loss. Implementing the PSD pipeline in SPADE allowed parameters such as the integration time and FFT point size to be changed in real-time without compromising the flow of data. 

Streams facilitates both implicit and explicit parallelization. Implicit parallelization is achieved by fusing operators into processing elements, and explicit parallelization by deployment of processing elements to many physical nodes. Furthermore Streams goes beyond conventional parallelization middleware and frameworks such as MPI (Message Passage Interface) and OpenMP by allowing dynamic operator fusing and processing element deployment to physical nodes during runtime. This degree of dynamic operation enables Streams to provide on demand scalability to increasing data loads and computations.

In this work we mainly focus on reviewing InfoSphere Streams and its potential use for Radio Astronomy. In our opinion the Streams approach has shown positive results to warrant further research and serious consideration for managing DPDM aspects of large antennae arrays. Further research is required to conduct more formal and specific comparative analysis between Streams and other middleware such as MPI and ICE (Internet Connection Engine). Another interesting area that requires more rigorous investigation is Streams scaling capabilities using a larger cluster of x86 nodes as well as combining other accelerators such as GPUs, Intel MIC (Many Integrated Core) architecture and FPGAs.


%
\begin{acknowledgement}
The first author would like to thank the New Zealand government's Tertiary Education Commission Build IT fund for funding this research and the IBM cooperation for providing the computing hardware under the shared university research grant.
\end{acknowledgement}

\section*{Appendix}
\addcontentsline{toc}{section}{Appendix}
The following SPADE code listing shows an implementation of an autocorrelation spectrometer application. Lines 4-7 define the virtual streams used by the applications stream operators. In lines 8-11 is a user defined source operator responsible for receiving network digitized raw antenna data using TCP. In lines 12-15 a split operator is used to distribute data to the PowerCell accelerators (for simplicity this listing uses two accelerators). In lines 17-20 an MTUDOP is used to send time series data and receive frequency domain data to and from the PowerCell accelerators. The frequency domain data received from a particular accelerator is then autocorrelated by a functor operator in lines 21-24. Since the accelerators are given a chunk containing multiple FFTs lines 25-41 integrate this chunk. In lines 43-46 an aggregate operator is used to integrate averaged chunks. In this particular listing the aggregate operator sends a result to the sink operator in line 47 every time it integrates 523 averaged PSD chunks.
\begin{lstlisting}
[Application]
	AutoCorrelator
[Program]
	vstream RawData(data : ShortList)
	vstream RawDataChunk(cellID: Integer, schemaFor(RawData))
	vstream ChannelData(real : FloatList, imag : FloatList)
	vstream PowerSpectrumData(psd : FloatList)
	stream Antenna(schemaFor(RawDataChunk))
		:= Source() ["stcp://thishost:9932/",
		udfBinFormat="AntennaParser",
		blockSize=8*1024] {}
	for_begin @Blade_ID 1 to 2
		stream QS22@Blade_ID(schemaFor(Antenna))
	for_end
		:= Split(Antenna)[cellID]{}
	for_begin @Blade_ID 1 to 2
		stream FFT@Blade_ID(schemaFor(ChannelData)) :=
			MTUdop(QS22@Blade_ID)["MT_QS22_FFT"] {
				switch1="@Blade_ID", switch2="9933", switch3="9934"
			}
		stream PSD@Blade_ID(schemaFor(PowerSpectrumData)) 
			:= Functor(FFT@Blade_ID) [] {
				psd := apply(pow, real, 2.0) .+ apply(pow, imag, 2.0)
			}
		stream IntegrateChunk@Blade_ID(schemaFor(
			PowerSpectrumData)) := Functor(PSD@Blade_ID)
		<
			Integer $count := 1;
			FloatList $average := makeFloatList();
		>
		<
			$average := slice(psd, 0, 4096);
			while($count < 512) {
				$average := $average .+ slice(psd, $count * 4096, 4096);
				$count := $count + 1;
			}					
			$average := $average ./ 512.0;
			$count := 1;
		>
		[true]
		{psd := $average}
	for_end
	stream Integrate(schemaFor(PowerSpectrumData)) := 
		Aggregate(IntegrateChunk1, IntegrateChunk2 <count(523)>) [] {
			psd := Avg(psd)
		}
	Nil := Sink(Integrate)["file:///../data/spectrum.bin", nodelays, udfBinFormat="DataFormatter"] {}
\end{lstlisting}
%
%
%
%

\end{document}